\documentclass[aps,prb,onecolumn,groupaddress,showpacs,11pt]{revtex4-1}
\usepackage{graphicx}
\usepackage{color}

\begin{document}

\title{Classical and quantum ordering of protons in cold solid hydrogen
  under megabar pressures}

\author{Xin-Zheng~Li$^{1,2}$}
\author{Brent~Walker$^{1}$}
\author{Matthew~I.~J.~Probert$^4$}
\author{Chris~J.~Pickard$^3$}
\author{Richard~J.~Needs$^5$}
\author{Angelos~Michaelides$^1$}
\email{angelos.michaelides@ucl.ac.uk}

\affiliation{
  $^1$London Centre for Nanotechnology and Department of Chemistry, University College London, London WC1E 6BT, U.K. \\
  $^2$School of Physics, Peking University, Beijing 100871, P. R. China \\
  $^3$Department of Physics and Astronomy, University College London, London WC1E 6BT, U.K. \\
  $^4$Department of Physics, University of York, York YO10 5DD, U.K. \\
  $^5$Theory of Condensed Matter Group, Cavendish Laboratory,
  University of Cambridge, J. J. Thomson Avenue, Cambridge CB3 0HE, U.K.  
  }
\date{\today}

\begin{abstract}  
  A combination of state-of-the-art theoretical methods has been used
  to obtain an atomic-level picture of classical and quantum ordering
  of protons in cold high-pressure solid hydrogen.  
  We focus mostly on phases II and III of hydrogen, exploring the
  effects of quantum nuclear motion on certain features of these
  phases (through a number of \textit{ab initio} path integral
  molecular dynamics (PIMD) simulations at particular points on the
  phase diagram).
  We also examine the importance of van der Waals forces in this
  system by performing calculations using the optB88-vdW density
  functional, which accounts for non-local correlations.
  Our calculations reveal that the transition between phases I and
  II is strongly quantum in nature, resulting from a competition
  between anisotropic inter-molecular interactions that restrict
  molecular rotation and thermal plus quantum fluctuations of the
  nuclear positions that facilitate it. 
  The transition from phase II to III is more classical
  because quantum nuclear motion plays only a secondary role and the
  transition is determined primarily by the underlying potential
  energy surface.  
  A structure of $P2_1/c$ symmetry with 24 atoms in
  the primitive unit cell is found to be stable when anharmonic
  quantum nuclear vibrational motion is included at finite
  temperatures using the PIMD method.  This structure gives a good
  account of the infra-red (IR) and Raman vibron frequencies of phase
  II.  We find additional support for a $C2/c$ structure as a strong
  candidate for phase III, since it remains transparent up to 300 GPa,
  even when quantum nuclear effects are included.  
  Finally, we find that accounting for van der Waals forces
  improves the agreement between experiment and theory for the parts
  of the phase diagram considered, when compared to previous work which
  employed the widely-used Perdew-Burke-Ernzerhof (PBE)
  exchange-correlation functional.
\end{abstract}
\maketitle

\section{Introduction}
Hydrogen (H) is the most abundant of the elements and, having one
electron, it can form only a single strong covalent bond.
As a result, pure H is expected to remain molecular to very high
pressures.
Several solid phases of H have been observed; phase I is a quantum
crystal of rotating H$_2$ molecules on a hexagonal close packed (hcp)
lattice but, despite decades of study, the arrangements of the
molecules in phases II and III are unknown.~\cite{Mao_1994,
  Ceperley_2012} 
Evidence has recently been found for a
room-temperature phase IV of solid hydrogen,~\cite{Eremets_2011,
  Howie_2012, Pickard_hydrogen_2012,Pickard_hydrogen_2012_Erratum}
although here we study only low temperatures of $T \leq 150$ K.

Determining the atomic structures of the different phases presents
a formidable challenge to experiment.
Direct experimental structure
determinations for solid hydrogen are hampered by the weak scattering
of X-rays by H and the difficulties of combining high-pressure static
diamond anvil cell techniques with such measurements.
To date only a single X-ray study covering the three low-temperature
phases has been reported.~\cite{Akahama_2010}
Although this study provides
constraints on the experimental structures and phase transitions, the
strongest experimental evidence comes from IR, Raman, and optical
measurements which provide valuable but indirect structural
information.~\cite{Mao_1994}
A key experimental observation is that the boundary between phases I
and II depends strongly on the isotope.~\cite{Mao_1994, Mazin_1997,
  Cui_1994} At low temperatures, the I/II transition occurs at 28 to
40~GPa~(0.28 to 0.4~megabar) in D$_2$,~\cite{Cui_1994, Silvera_1981}
50 to 69~GPa in HD,~\cite{Moshary_1993} and 90 to 110~GPa in
H$_2$.~\cite{Mazin_1997, Lorenzana_1990} In contrast, the boundary
between phases II and III, which occurs at
$\sim$160~GPa,~\cite{Goncharov_2011} is only weakly dependent on the
isotope.
Likewise, the transition from phase I to II is strongly temperature
dependent, whereas the II/III transition is essentially independent of
temperature.
Upon passing from phase I to II the vibrational roton modes undergo
substantial changes, indicating that the molecules change from a
rotationally free state to a rotationally restricted one.
The IR activity increases dramatically at the II/III transition and
the IR and Raman vibron frequencies soften by about
80~cm$^{-1}$.~\cite{Akahama_2010, Loubeyre_2002}
Phase III remains transparent up to 300 GPa but becomes dark by about
320 GPa.~\cite{Loubeyre_2002}

The theoretical characterisation of high pressure solid hydrogen is
also a major challenge.
The development of a successful model
must start with an extensive exploration of structural phase space in
order to identify candidate structures for the different phases.~\cite{Pickard_2007,Tse_2008}
In addition, the energetics of the structures must be accurately
accounted for, and the optical properties must be well-described,
particularly at the high pressures at which metallisation may occur.
Given the low mass of hydrogen, quantum nuclear effects including
zero-point~(ZP) motion should also be accounted
for.~\cite{Kohanoff_1997,Biermann_1998a, Biermann_1998b, Kitamura_2000}
No current theoretical method is capable of giving a satisfactory
account of all of these, and none of the excellent theoretical papers on this system have 
addressed all of these issues simultaneously.  In this work we have 
addressed these points using a combination of
state-of-the-art methods.
%
We have used the results from extensive
density-functional-theory~(DFT) based searches of the potential energy
surface~(PES) \cite{Pickard_2007, Pickard_2009}, so that our
simulations start from the most stable structures identified to date.
We have used the \textit{ab initio} path-integral molecular dynamics
(PIMD) method \cite{Marx_1994, Marx_1996, Tuckerman_1996} to account
for quantum nuclear motion at finite temperatures.
The equilibrated structures obtained from the PIMD calculations are
used to calculate optical properties using the $GW$ many-body
perturbation theory approach, and a hybrid exchange-correlation
density functional.~\cite{Johnson_2000}
In the PIMD simulations we use a 144-atom supercell together with the
constant-pressure, constant-temperature~(NPT)
ensemble.~\cite{Martyna_1999}
This ensemble allows relaxation of the cell shape and size, and
reduces the bias towards particular structures which arises from using
small fixed cells.
The use of large variable cells and lower-enthalpy static-lattice
structures as the starting points of our simulations are important
advances over previous PIMD studies of solid H, where either ``quantum
localisation'' or metallic structures at too low pressures were
predicted.~\cite{Biermann_1998a, Biermann_1998b, Kitamura_2000}
As we show below, the broad range of
methods employed here provides an atomic-level picture of the
classical and quantum ordering of protons in low-temperature solid
hydrogen that is consistent with the key experimental results.

\section{Computational details}

The quantum behaviour of the nuclei is described within the PIMD
method by a ``ring polymer'' consisting of beads connected by
springs. \cite{Marx_1994, Marx_1996, Tuckerman_1996} Nuclear exchange
effects are neglected in our calculations and we assume an adiabatic
decoupling of the electron and nuclear motions.  PIMD is a finite
temperature method and the number of beads required to give an
accurate sampling of the imaginary time path decreases with increasing
temperature. We have used 64~beads per nucleus for the simulations at
50 and 100~K, and 16~beads at 150~K.
We have performed NPT PIMD simulations at 80 and 200~GPa.
The simulations at 80~GPa were run for 9,000 steps using a time step
of 0.25~fs, and the final 8,000 steps were included in the data
analysis.  Those at 200~GPa were run for 20,000 steps with a time step
of 0.125~fs, and data from the final 15,000 steps were included in the
analysis.
Convergence tests were performed on the number of beads used, the
simulation length, and the sampling intervals, and we also tested the
validity of neglecting the nuclear exchange in the PIMD calculations,
as reported in the supporting information.
The DFT-based classical molecular dynamics~(MD) and PIMD simulations
were of the Born-Oppenheimer type, \textit{i.e.}, the electronic
density was optimised self-consistently to the ground state at each
time step.
The electronic states were sampled on a k-point grid of spacing
$0.05\times2\pi/\text{\AA}$ in all of the calculations.
The MD and PIMD simulations were carried out using the CASTEP code
\cite{CASTEP} with ultra-soft pseudopotentials \cite{USP} and a 350 eV
plane-wave cut-off.  The IR and Raman spectra were calculated within
density-functional perturbation theory \cite{DFPT}, using the CASTEP
code, norm-conserving pseudopotentials \cite{Norm1, Norm2} and a
plane-wave cut-off energy of 800 eV.  All of the CASTEP calculations
were performed with the Perdew-Burke-Ernzerhof~(PBE) exchange-correlation
functional~\cite{PBE-GGA}. The vdW-DF (optB88-vdW)~\cite{Klimes_2010},
$GW$, and hybrid functional calculations were carried out using the
VASP~\cite{VASP3,Klimes_2011} code with PAW
potentials~\cite{PAWPP1,PAWPP2} and a 600 eV plane-wave cut-off. The
X-ray diffraction simulations were performed using the Powder Cell
program.~\cite{POWDERCELL}

\section{Results and Discussions}

\subsection{Accuracy of PBE}

\begin{figure}
\includegraphics[width=4.0in,clip]{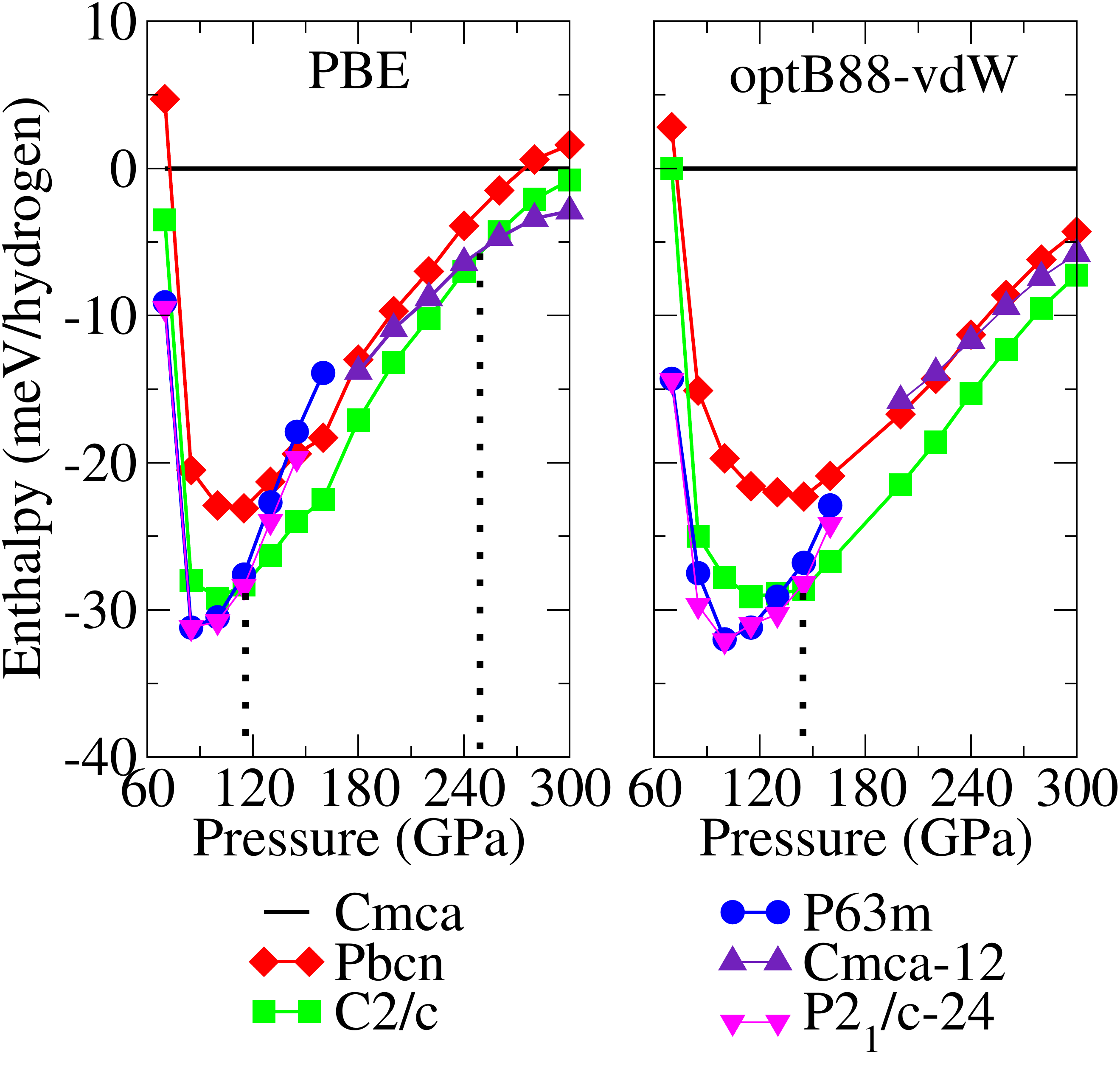}
\caption{\label{figure1} Enthalpy as a function of pressure relative
  to the $Cmca$ structure \cite{Johnson_2000} at megabar pressures.
  The results shown in the left panel were calculated using the PBE
  functional and those in the right panel were obtained using the
  optB88-vdW functional, which accounts for van der Waals (vdW)
  interactions within the vdW-DF scheme.  The $Cmca$-12 molecular
  structure is described in Ref.\ \onlinecite{Pickard_2007}. }
\end{figure}

Pickard and Needs performed extensive
explorations of the structural phase space of hydrogen, and a number
of low-enthalpy structures were identified at pressures of about 100
GPa where phase II is stable.~\cite{Pickard_2007, Pickard_2009}
These structures correspond to minima
in the static lattice enthalpy, and they do not include any effects from vibrations.
The molecular centres in these structures lie very close to a hcp
lattice, as is generally thought to occur in phase II.
In this study we have used the $P2_1/c$-24 structure
\cite{Pickard_2009} as a model for phase II.
The $P2_1/c$-24 structure has 24 atoms in the primitive unit cell
\cite{Pickard_2009}, and it is a little more stable than the structure
of the same symmetry reported in Ref.\ \onlinecite{Johnson_2000},
which has an 8-atom primitive unit cell.
At higher pressures these hcp-based structures become unstable with
respect to a layered structure of $C2/c$ symmetry with 24 atoms in the
primitive cell.  A related and slightly less stable structure, with
the same space group but a smaller unit cell, was found by Tse
\textit{et al.}~\cite{Tse_2008}.
DFT calculations using the PBE exchange-correlation functional show that $C2/c$
has the lowest static-lattice enthalpy among the plausible candidates
for phase III in most of the pressure range in which it has been
observed, see Fig.\ \ref{figure1} and Refs.\ \onlinecite{Pickard_2007}
and \onlinecite{Pickard_2009}.
Although PBE has been successfully applied in many high pressure
studies, it does not account for van der Waals~(vdW) interactions,
which are often important in binding molecular crystals.
We have therefore tested the effects of including vdW forces on the
relative enthalpies and phase boundaries using the optB88-vdW
functional \cite{Klimes_2010} within the vdW-DF
scheme.~\cite{Dion_2004}
Fig.\ \ref{figure1} shows that the relative stabilities of the
structures are generally not altered substantially.
However, the transition pressure between phases II and III (taking the
$P2_1/c$-24 structure as our model for phase II and $C2/c$ for phase
III) is increased from $\sim$120~GPa with PBE to $\sim$150~GPa with
optB88-vdW, in better agreement with experiment. \cite{Goncharov_2011}
As can be seen in Fig.\
\ref{figure1}, static lattice PBE calculations indicate that $Cmca$
\cite{Johnson_2000} is less favourable than $C2/c$ over the pressure
range in which phase III is observed.  However, including harmonic ZP
motion changes the picture significantly
\cite{Pickard_hydrogen_2012,Pickard_hydrogen_2012_Erratum}, and the
softer $Cmca$ structure becomes the most energetically favourable
above about 225 GPa.
The $Cmca$ structure is, however, metallic and its vibronic
frequencies are considerably lower than those of phase III, and
therefore $Cmca$ is not a plausible model for phase III.
As can be seen from Fig.\ \ref{figure1}, using the optB88-vdW
functional~\cite{Klimes_2010} instead of PBE leads to an increase in
the enthalpies with respect to $C2/c$ of about 6 meV per proton for
the weakly molecular $Cmca$ structure and about 3 meV per proton for
the strongly molecular $Cmca$-12 structure.  When harmonic ZP motion
corrections are added to the static lattice data of Fig.\
\ref{figure1} we find that $C2/c$ remains the most stable phase up to
about 300 GPa.  
Using the optB88-vdW functional therefore removes a
major conflict with experiment in the PBE phase diagram reported in
Refs.\ \onlinecite{Pickard_hydrogen_2012,Pickard_hydrogen_2012_Erratum}
and, overall, gives significantly better agreement with the
experimental phase diagram than the PBE functional.

\subsection{Phases I and II}

In order to explore phases I and II in more detail, we have carried
out an extensive series of \textit{ab initio} PIMD simulations.  The
first set of calculations was performed at 80 GPa, a pressure safely
within the experimental region of stability of phase II in solid D$_2$
and the region in which static DFT calculations predict phase II-like
structures to be stable.
PIMD simulations were performed on both the $P6_3/m$ and $P2_1/c$-24
structures since they are the two lowest-enthalpy static-lattice structures at
80 GPa.
A 144-atom supercell was used which is commensurate with the cells of
both the $P6_3/m$~\cite{Pickard_2007} and $P2_1/c$-24
structures.~\cite{Pickard_2009}
Using D at a temperature of 50 K and starting from the $P6_3/m$
structure, MD and PIMD simulations quickly showed a transformation
from the original structure to a new rotationally restricted state.
The centres of the molecules still form a hcp lattice but the
molecular orientations differ from the starting state and it is
difficult to identify a corresponding static structure.
This indicates that although $P6_3/m$ corresponds to a minimum in the
PES (with positive harmonic vibrational frequencies), it is so shallow
that thermal or quantum effects are sufficient to disrupt it.
Therefore $P6_3/m$ is not a plausible structure for phase II when
anharmonic nuclear motion is taken into account.

\begin{figure}
\includegraphics[width=3.0in]{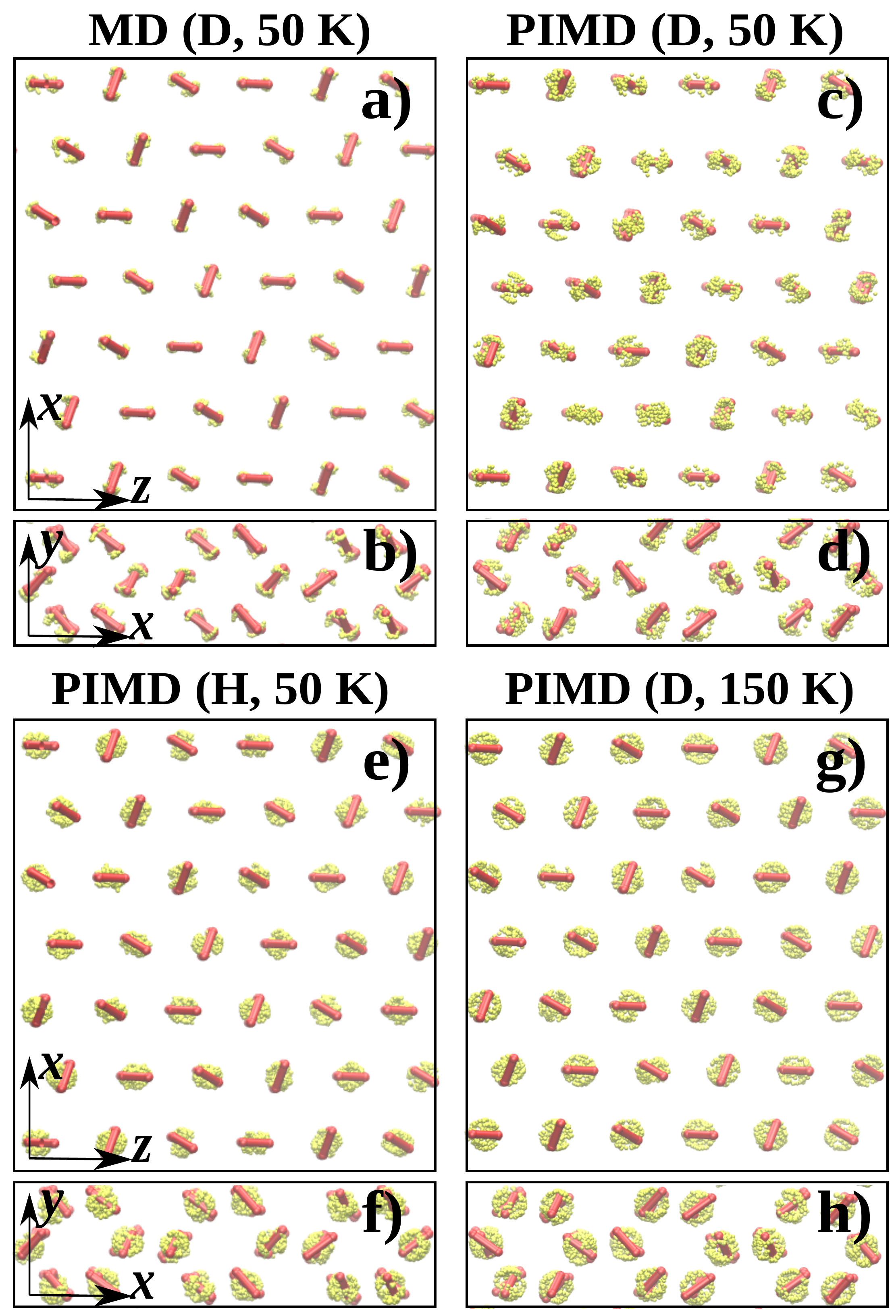}
\caption{\label{figure2} Trajectories of structures obtained from
  simulations with classical and quantum nuclei at 80 GPa starting
  from the $P2_1/c$-24 structure. Yellow balls show the representative
  configurations of the centroids throughout the course of the
  simulation. The red rods show the static (geometry-optimised)
  structure. A conventional hexagonal cell containing 144 atoms was
  used.  Panels a, c, e, and g show the $z-x$ plane and panels b, d,
  f, and h show the $x-y$ plane of the hcp lattice.  The four
  simulations are: 1) MD with classical nuclei at 50 K (panels a and
  b), 2) PIMD for D at 50 K (panels c and d), 3) PIMD for H at 50 K
  (panels e and f), and 4) PIMD for D at 150 K (panels g and h).  In
  the MD simulation, the anisotropic inter-molecular interaction
  outweighs the thermal and quantum nuclear fluctuations. Therefore,
  the molecular rotation is highly restricted.  The thermal plus
  quantum nuclear fluctuations outweigh the anisotropic
  inter-molecular interactions in the PIMD simulations of H at 50~K
  and D at 150~K. }
\end{figure}

\begin{figure}
\includegraphics[width=5.0in]{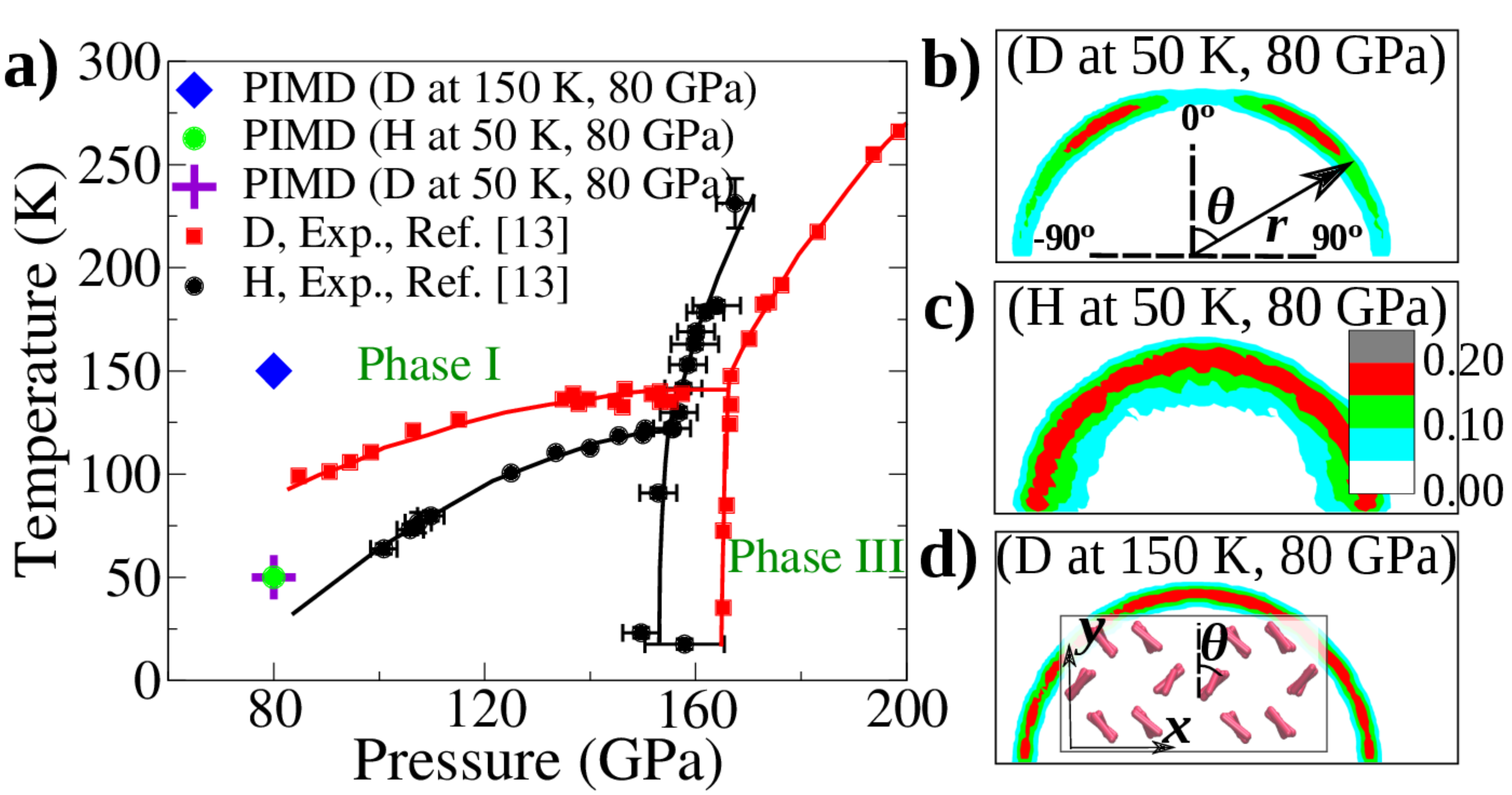}
\caption{\label{figure3} Phase diagram
  of solid H and D with the pressure-temperature coordinates of three
  key PIMD simulations indicated at 80~GPa (in panel a), and
  probability distribution functions for these simulations (panels b,
  c, and d). The three PIMD simulations are: 1) D at 50~K, 2) H at
  50~K, and 3) D at 150~K. D at 50~K has the least thermal plus
  quantum nuclear fluctuations, while H at 50~K and D at 150~K have
  larger thermal plus quantum fluctuations. The probability is plotted as a
  function of $r$ (molecular bond length) and $\theta$ (angle between
  the projection of the molecules on the $x-y$ plane and the $y$
  axis). The ground state structure has molecules aligned between
  30$^\circ$ and 50$^\circ$ and between $-$30$^\circ$ and
  $-$50$^\circ$ (inset in panel d and Fig.~S1 of the supporting
  information). The molecules of D at 50 K (panel b) are still
  rotationally restricted (phase~II). Replacing D by H (panel c) or
  elevating the temperature to 150 K (panel d) leads to free rotation
  (phase~I). Thermal plus quantum fluctuations compete with the
  anisotropic inter-molecular interactions in the region of the
  transition from phases I to II. }
\end{figure}

Starting from the $P2_1/c$-24 structure, we performed a classical MD
simulation at 50 K, PIMD runs for D and H at 50 K, and a PIMD run for
D at 150 K.
Some of the key results from these simulations are reported in Fig.\
\ref{figure2}, where representative configurations of the centroid
trajectories are plotted.
We find that the orientational ordering of the starting structure is
maintained in the classical MD simulation (Fig.\ \ref{figure2}(a) and
(b)).  Similarly, the initial orientational ordering is rather well
preserved in the PIMD simulation with D (Fig.\ \ref{figure2} (c) and
(d)).
Under the same conditions, but switching to the lighter H atom, we
find that the orientational ordering is almost completely lost, as
reflected by the spherical distributions seen in Fig.\
\ref{figure2}(e) and (f).
Returning to the heavier D atom but increasing the temperature to 150
K, we again find a spherical trajectory distribution on a hcp
lattice. These qualitative interpretations are supported by a detailed
analysis of the probability distributions associated with the
rotational freedom of the molecules.
We calculate the angle $\theta$ between the projection of the
molecules on the $x-y$ plane and the $y$ axis (inset of Fig.\
\ref{figure3}~(d)), and plot probability distributions of $\theta$ and
the molecular bond length ($r$) for each simulation.
The geometry-optimised $P2_1/c$-24
structure has molecules aligned between 30$^\circ$ and 50$^\circ$ and
between $-$30$^\circ$ and $-$50$^\circ$.
Two rather broad peaks are seen in the PIMD simulations for D at 50 K
at about $\pm$40$^\circ$ (Fig.\ \ref{figure3}~(b)), which are
consistent with restricted molecular rotation.
On the other hand, either replacing D by H or raising the temperature
to 150 K destroys the peak structure and leads to a probability
distribution characteristic of a freely rotating molecular state
(Fig.\ \ref{figure3} (c) and (d)).
These results show that the degree of orientational order is governed
by a competition between the corrugation of the underlying PES, which
favours rotational restriction, and the thermal and quantum nuclear
fluctuations, which oppose it.  We find no evidence for the ``quantum
localisation'' reported in previous PIMD
simulations.~\cite{Kitamura_2000}

Our results explain in a simple manner the positive slope
($\text{d}P/\text{d}T$) of the I/II coexistence line as well as the
isotope-dependence of the transition pressure.
The corrugation of the underlying PES
increases with pressure and, as the pressure increases, higher
temperatures are required to overcome the corrugation.
The I/II transition for hydrogen occurs
at a higher pressure than in deuterium for a similar reason; quantum
effects are greater for hydrogen and so the corrugation of the
underlying PES is more easily overcome by hydrogen than by deuterium.
These results are consistent with the traditional understanding of the
quantum nature of phase I. \cite{Mao_1994, Mazin_1997}

\subsection{Phase III}

\begin{figure}
\includegraphics[width=5.0in]{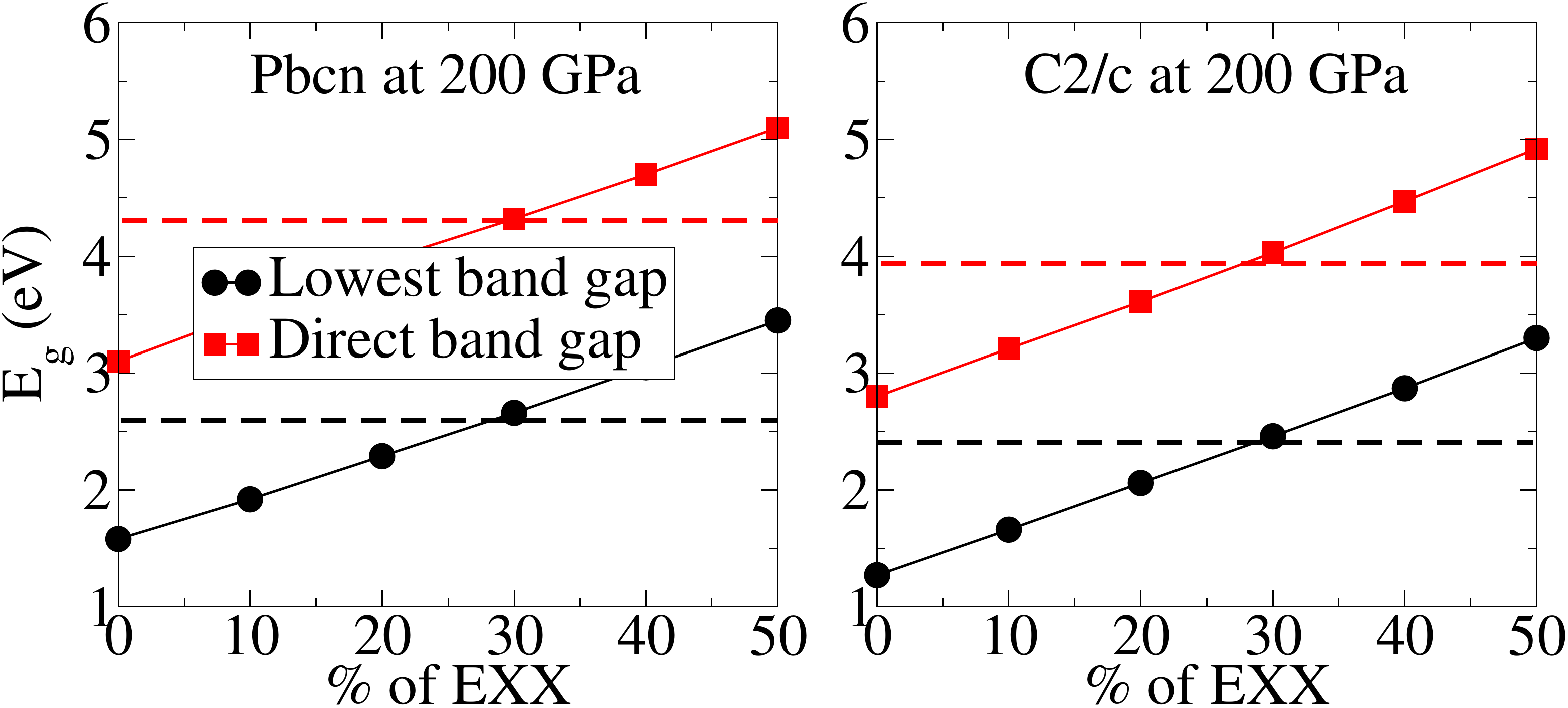}
\caption{\label{figure4} Variation of the lowest and direct
  band gaps with the percentage of exact-exchange (EXX) in the hybrid
  functional calculations for the static $Pbcn$ (left panel) and
  $C2/c$ (right panel) structures at 200~GPa.  The black (red) dashed
  line indicates the LDA based $G_0W_0$ results for the lowest
  (direct) band gap. The $G_0W_0$ results are converged to within
  0.01~eV with respect to the number of unoccupied states.  Using 30\%
  exact-exchange reproduces the $G_0W_0$ band gaps to within
  $\sim$0.1~eV.}
\end{figure}

\begin{figure}
\includegraphics[angle=-90,width=5.0in]{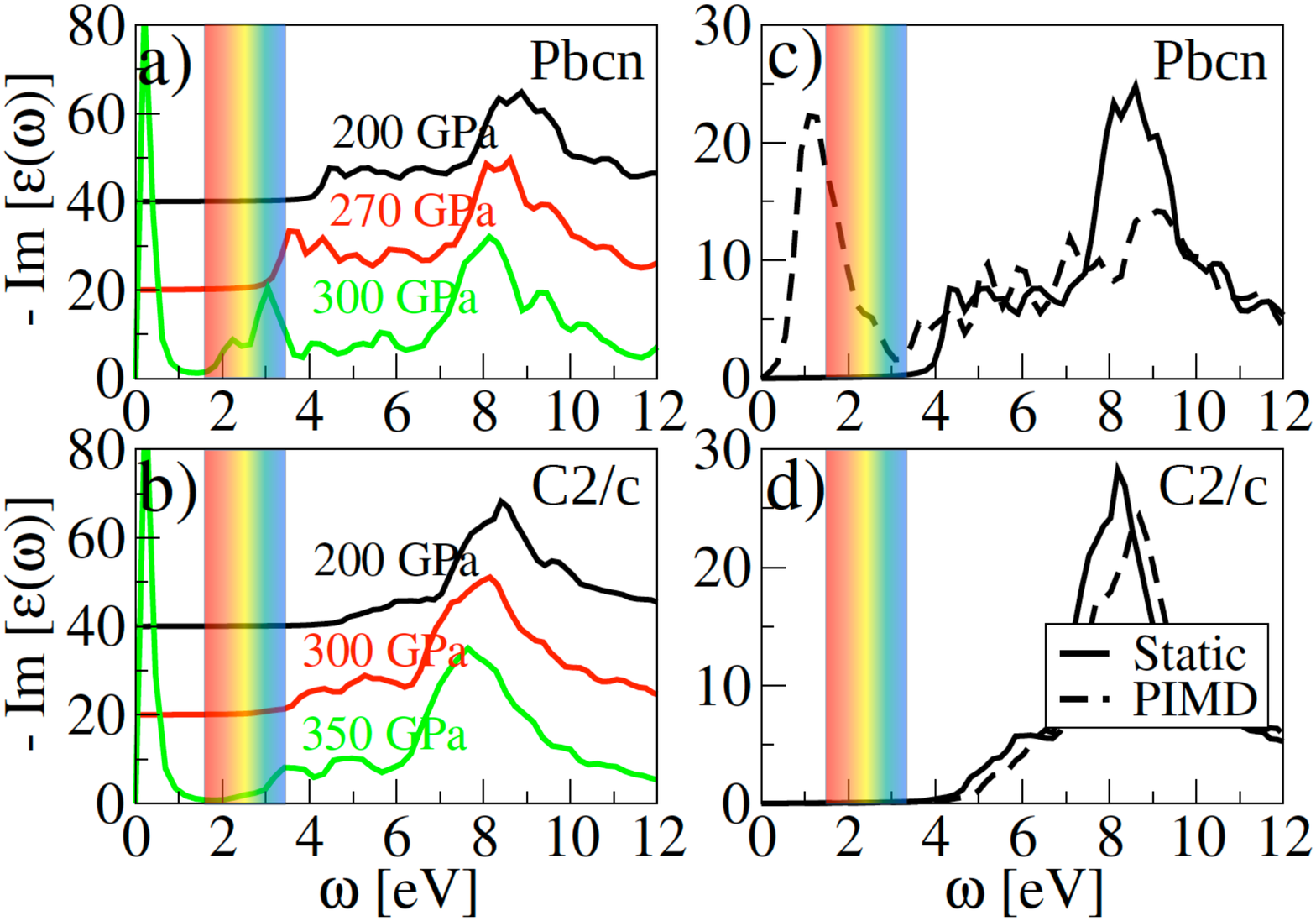}
\caption{\label{figure5} Absorption spectra of hydrogen at 200~GPa.
  Panels~(a) and (b) show absorption spectra calculated using 30\%
  exact exchange at different pressures for (a) $Pbcn$, and (b)
  $C2/c$, using static geometry-optimised structures.  Panels (c)
  ($Pbcn$) and (d) ($C2/c$) compare the 200 GPa absorption spectra
  obtained from static geometry-optimised structures (solid lines)
  with those obtained from the PIMD simulations (dashed lines). The
  visible light region is shown by the rainbow in each panel.  Using
  the PIMD structure renders $Pbcn$ opaque at 200~GPa, but has little
  effect on $C2/c$.}
\end{figure}

DFT structure searches \cite{Pickard_airss_2011} have so far provided
a single best candidate for phase III: a layered $C2/c$ structure with
significant deviations from close packing within the
layers.~\cite{Pickard_2007}
Other competitive structures were found, including an intriguing
structure of $Pbcn$ symmetry which consists of alternate layers of
strongly bonded and more weakly bonded molecules (see Fig.\ 3 of the
supplementary information for Ref.\ \onlinecite{Pickard_2007}).
$Pbcn$ has a lower ZP enthalpy than structures consisting entirely of
strongly bonded molecules, and it might be favoured when ZP motion is
taken into account. We therefore performed a number of simulations for
both $Pbcn$ and $C2/c$.  In particular, we focused on understanding
whether either of these structures is consistent with the experimental
observation that phase III is transparent up to 300 GPa, but becomes
opaque by 320~GPa.~\cite{Loubeyre_2002} 
To this end we calculated the band gap energies of the $Pbcn$ and
$C2/c$ structures using a many-body perturbation theory approach and
the $GW$ approximation to the self energy, and PBE0-like density
functionals with different percentages of ``exact-exchange''
(EXX).~\cite{PBE0} We found that using a hybrid density functional
including 30\% EXX reproduced the direct and lowest band gaps of the
$Pbcn$ and $C2/c$ structures obtained within $GW$ theory (Fig.\
\ref{figure4}).  This hybrid density functional was subsequently used
to calculate the optical absorption spectra of the static lattice
$Pbcn$ and $C2/c$ structures at pressures between 200 and 350 GPa, see
Fig.\ \ref{figure5}(a) and (b).  The peaks
in the absorption spectra correspond to strong inter-band transitions,
and a non-zero imaginary part of the macroscopic dielectric function
in the visible light range indicates that the material is to some
extent opaque.
The evolution of the absorption spectra shows that $Pbcn$ becomes
opaque between 270 and 300~GPa (Fig.\ \ref{figure5}a), while $C2/c$
turns opaque between 300 and 350~GPa (Fig.\ \ref{figure5}b). 
Considering the limited accuracy of the static pressure, the
static-lattice absorption spectra of both $Pbcn$ and $C2/c$ are in
qualitative agreement with the experimental observations.

We now consider the changes in the absorption spectra induced by
introducing thermal and quantum nuclear fluctuations.
To this end we have performed PIMD simulations for H using 48-atom
cells, starting from the $Pbcn$ and $C2/c$ structures, at 100~K and
200~GPa, a pressure well within the stability range of phase III.
These reveal that the $Pbcn$ structure maintains its original symmetry
but the weak molecular layers evolve to atomic-like layers in which
the nearest neighbour distances are approximately equal.
When using the new PIMD structure to calculate the absorption spectra
we find that $Pbcn$ is already opaque at 200 GPa (note the absorption
peak within the visible light region in Fig.\ \ref{figure5}(c)).
This is inconsistent with experiments on phase III which find that it
remains transparent up to $\sim$300 GPa.
However, in the PIMD simulations for the $C2/c$ phase, the structure
and consequently the absorption spectrum are only slightly altered
compared with the original geometry-optimised structure (Fig.\
\ref{figure5}(d)).
Therefore, unlike $Pbcn$, $C2/c$ remains transparent up to pressures
as high as 300~GPa when thermal and quantum fluctuations are taken
into account.
Zha \textit{et al.}\ \cite{zha} recently reported synchrotron IR
measurements for dense hydrogen, finding that the IR vibron of phase
III persists up to at least 360~GPa.  This behaviour is also
reasonably consistent with the $C2/c$ structure.

\subsection{Discussion and Conclusions}
From the previous discussion it is clear that our most promising
candidate structure for phase II is $P2_1/c$-24, but we have
insufficient evidence to identify it as the correct structure.
The $P2_1/c$-24 structure is
monoclinic, while experiments have suggested that phase II is
hexagonal. \cite{Goncharenko_2005}
Experiments on D$_2$ have suggested
that phase II is incommensurate \cite{Goncharenko_2005}, and in a
theoretical study \cite{Biermann_1998a} it was speculated that phase
II of H$_2$ might be ``diffuse'', meaning that it does not correspond
to a single classical structure.
We assign space group symmetries to the
static lattice structures, not to the structures found in the PIMD
simulations which include nuclear motion.  It is possible that the
nuclear motion might reduce lattice distortions, similar to the way in
which heating can lead to more symmetric structures, although such an
effect was not apparent in our PIMD simulations.
The differences between the results
obtained with the PBE and optB88-vdW functionals indicate some
sensitivity to the functional used, which must temper our confidence
in the accuracy of DFT results for this system.
In addition, our simulations do not account for nuclear exchange
effects such as those that occur in \textit{ortho} and \textit{para}
molecules, which are known to be important in a quantitative
description of phase I and its transition to phase II~\cite{igarashi,
  goncharov, freiman1, freiman2}.
Notwithstanding these limitations, it is clear that quantum nuclear
effects and molecular rotations play a very important role in the
transition between phases I and II and the transition is strongly
quantum in nature even when nuclear exchange is neglected.  

\begin{figure}
\includegraphics[width=5.0in]{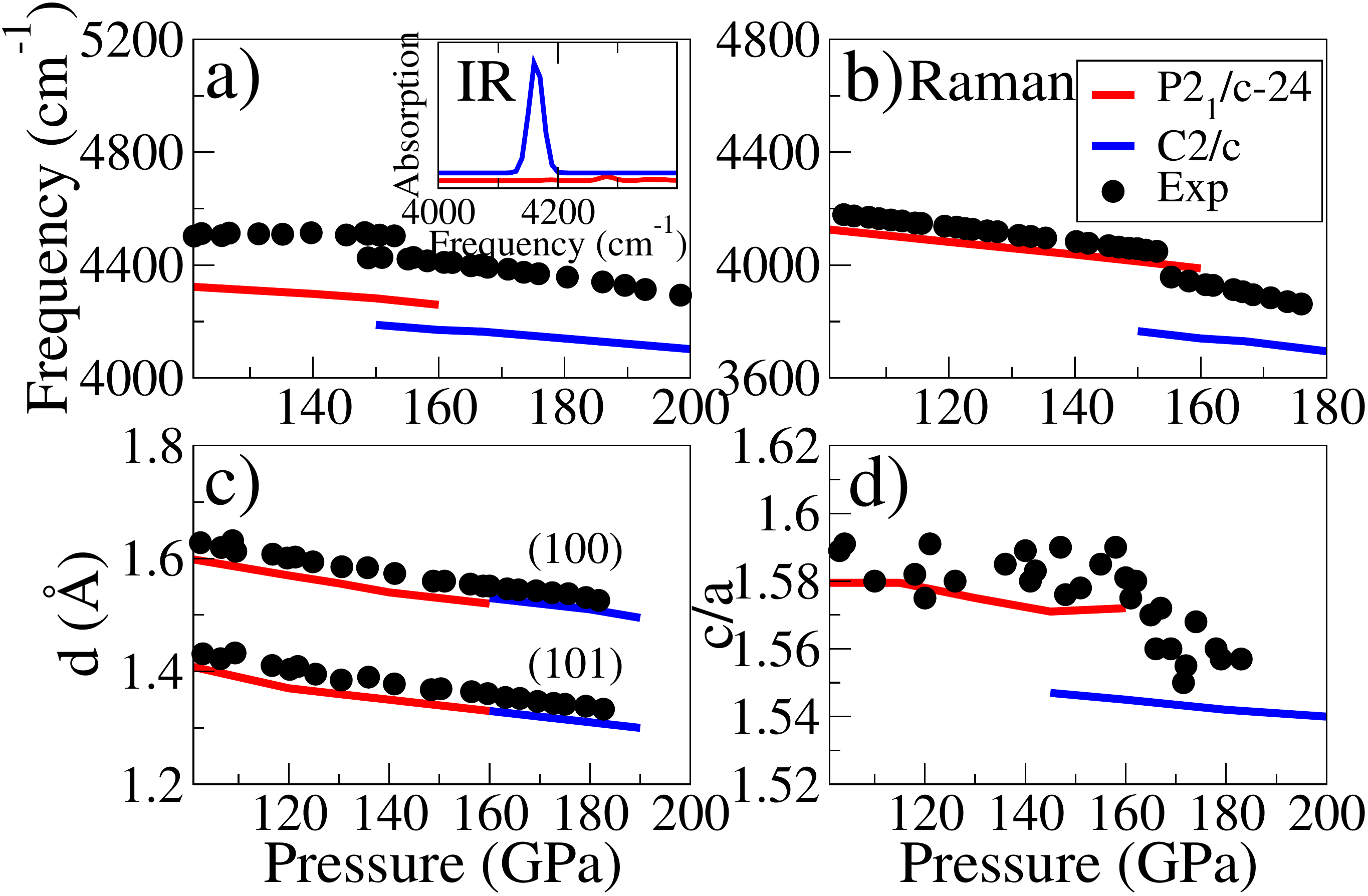}
\caption{\label{figure6} Comparison of calculated and experimental
  properties of phases II and III, assuming the $P2_1/c$-24 structure
  for phase II and $C2/c$ for phase III.  Variation with pressure of
  the IR (a) and Raman (b) vibron frequencies.  The inset in (a) shows
  the dramatic difference in IR intensity between $P2_1/c$-24 (phase
  II) and $C2/c$ (phase III).  The $C2/c$ spectrum is slightly shifted
  to distinguish it from that of $P2_1/c$-24. (c) Variation with
  pressure of the (100) and (101) X-ray diffraction peak
  positions. (d) Variation with pressure of the $c/a$ ratio.
  Experimental data is indicated by circles; the data in (a) and (b)
  were taken from Ref.\ \onlinecite{hanfland_1993} and those in (c)
  and (d) were taken from Ref.\ \onlinecite{Akahama_2010}.}
\end{figure}

As the pressure increases the molecules rearrange with a significant
distortion from close packing, so that the molecules can avoid one
another.  The difference in the thermal plus quantum motion of the two
phases plays only a secondary role and the II/III transition is
therefore only weakly dependent on the isotope and temperature.
Based on the observation of persistent (100) and (101) X-ray
diffraction peaks up to 180~GPa, Akahama \textit{et al.}\ have
concluded that any distortion from hcp packing in phase III is
small.~\cite{Akahama_2010} This appears to rule out the $C2/c$
structure.  However, we have calculated X-ray diffraction data
assuming $P2_1/c$-24 for phase II and $C2/c$ for phase III, finding
the positions of the (100) and (101) peaks in the hcp lattice to be in
very good agreement with experiment over a wide pressure
range,~\cite{Akahama_2010} see Fig.\ \ref{figure6}(c).  We have also
found the variation with pressure of the $c/a$ ratios of $P2_1/c$-24
and $C2/c$ to be in good agreement with experiment, including the
discontinuous drop at the II/III transition (Fig.\
\ref{figure6}(d)). \cite{Akahama_2010}
The variations with pressure of the IR
and Raman frequencies are well accounted for by the $P2_1/c$-24 and
$C2/c$ structures, see Fig.\ \ref{figure6}(a) and (b). 
The two ends of a molecule in the $C2/c$ structure
have different environments, so the molecules have dipole moments and
the structure shows intense IR vibron activity (Fig.\ \ref{figure6}(a)
inset), as found in experiments on phase III.

In conclusion, we have studied solid hydrogen at megabar pressures
using a variety of \textit{ab initio} techniques.
The transition from phase I to II is governed by a competition between
the corrugation of the underlying PES, which restricts molecular
rotation, and the thermal and quantum nuclear fluctuations, which
facilitate it.
At very low pressures and temperatures the
ZP motion washes out the corrugation of the PES and leads to the
freely-rotating phase I for both isotopes.
Including ZP motion destabilises the $P6_3/m$ structure, which is
therefore not a plausible candidate for phase II.
Our simulations favour the proposal that phase II is a hcp-based
rotationally restricted molecular phase with a large
cell.~\cite{Goncharenko_2005}
The IR and Raman active vibron frequencies of the $P2_1/c$-24
structure are in good agreement with the experimental data for phase
II.~\cite{hanfland_1993} We
also find that the (100) and (101) X-ray diffraction peak positions of
$P2_1/c$-24 and $C2/c$, and their $c/a$ ratios, are in good agreement
with experiment.~\cite{Akahama_2010}
The optB88-vdW~\cite{Klimes_2010} density functional gives a II/III
transition pressure in much closer agreement with experiment than PBE,
and it moves the $Cmca$ and $Cmca$-12 phases to higher enthalpies, so
that the $C2/c$ structure becomes the most stable phase up to about
300 GPa.  Using the optB88-vdW functional therefore removes a major
discrepancy between the theoretical and experimental phase diagrams.
We have found that $C2/c$ remains transparent up to 300 GPa, which
provides further support for it as a model for phase III.
Overall, our study provides evidence from a range of state-of-the-art
theoretical methods which supports the picture of orientational
ordering proposed by Mazin \textit{et al.}~\cite{Mazin_1997} and
others. 
We have provided an atomic-level picture of the evolution of solid
hydrogen under megabar pressures which satisfactorily explains some of
the experimental results, and can be tested by further experimental
and theoretical studies.

\begin{acknowledgments}
  This work was supported by the European Research Council and the
  EPSRC.  XZL thanks Felix Fernandez-Alonso for helpful discussions on
  X-ray scattering. We are grateful for computational resources
  supplied by the London Centre for Nanotechnology, UCL Research
  Computing, and the UK's national high performance computing service
  HECToR (for which access was obtained via the UKCP consortium,
  EP/F036884/1).
\end{acknowledgments}

\bibliography{ref}
\end{document}